\documentstyle[prl,aps,epsfig]{revtex}
\newcommand{\bmt}{\begin{mathletters}}
\newcommand{\emt}{\end{mathletters}}
\newcommand{\lb}[1]{\label{eq:#1}}
\newcommand{\eq}[1]{Eq.~(\ref{eq:#1})}
\newcommand{\ef}[1]{(\ref{eq:#1})}

\newcommand{\zx}{\zeta(x_1)}
\newcommand{\zxp}{\zeta(x_1^{\prime})}
\newcommand{\ew}{\epsilon (\omega )}

\newcommand{\aokp}{\alpha_0 (k')}
\newcommand{\aoqp}{\alpha_0 (q')}
\newcommand{\aok}{\alpha_0 (k)}

\newcommand{\ak}{\alpha(k)}
\newcommand{\alp}{\alpha (p)}
\newcommand{\aq}{\alpha (q)}
\newcommand{\nn}{\nonumber}
\newcommand{\aoq}{\alpha_0 (q)}
\newcommand{\ts}{\theta_s}
\renewcommand{\to}{\theta_0}

\newcommand{\iii}{\int^{\infty}_{-\infty}}
\newcommand{\bqe}{\begin{eqnarray}}
\newcommand{\eqe}{\end{eqnarray}}

\newcommand{\w}{\omega }
\newcommand{\e}{\epsilon}

\newcommand{\la}{\langle}
\newcommand{\ra}{\rangle}

\newcounter{tr}
\newcommand{\sctr}[1]{\setcounter{tr}{#1}}

\begin{document}

\title{The Angular Intensity Correlation Functions C$^{(1)}$ and 
C$^{(10)}$ for the Scattering of Light from Randomly Rough Dielectric 
and Metal Surfaces}
\author{Tamara A.\ Leskova$^a$, Ingve Simonsen$^{b,c}$, 
Alexei A.\  Maradudin$^c$}
\address{$^a$Institute of Spectroscopy, Russian Academy of Sciences,
Troitsk, Russia}
\address{$^b$Department of Physics, The Norwegian University of 
Science and Technology,  Trondheim, Norway}
\address{$^c$Department of Physics and Astronomy and Institute for Surface and
Interface Science,\\
University of California, Irvine, CA, USA}

\maketitle

\begin{abstract}
  We study the statistical properties of the scattering matrix
  $S(q|k)$ for the problem of the scattering of light from a randomly
  rough one-dimensional surface, defined by the equation $x_3 = \zx$,
  where the surface profile function $\zx$ constitutes a zero-mean,
  stationary, Gaussian random process, through the effects of $S(q|k)$
  on the angular intensity correlation function $C(q,k|q',k')$.  The
  existence of both the $C^{(1)}$ and $C^{(10)}$ correlation functions
  is consistent with the amplitude of the scattered field obeying
  complex Gaussian statistics in the limit of a long surface.  We show
  that the deviation of the statistics of the scattering matrix from
  circular Gaussian statistics and the $C^{(10)}$ correlation function
  are determined by exactly the same statistical moment.  As the
  random surface becomes rougher, the amplitude of the scattered field
  no longer obeys complex Gaussian statistics but obeys complex
  circular Gaussian statistics instead.  In this case, the $C^{(10)}$
  correlation function should vanish.  This result is confirmed by
  numerical simulation calculations.
\end{abstract}

\setcounter{equation}{0}
\section*{1.Introduction}
\setcounter{section}{1}

The scattering of light from randomly rough surfaces has attracted
attention over many years. The majority of the theoretical and
experimental studies of such scattering has been devoted to coherent
interference effects occuring in the multiple scattering of
electromagnetic waves from randomly rough surfaces and the related
backscattering enhancement phenomenon.

Recently, attention has begun to be directed toward theoretical [1-12]
and experimental [2,7,8,12,13] studies of multiple-scattering effects
on higher moments of the scattered field, in particular on angular
intensity correlation functions.  These correlation functions describe
how the speckle pattern, formed through the interference of randomly
scattered waves, changes when one or more parameters of the scattering
system are varied.

The interest in these correlations has been stimulated by the
expectation that, just as the inclusion of multiple-scattering
processes in the calculation of the angular dependence of the
intensity of the light that has been scattered incoherently from, or
incoherently through, a randomly rough surface, led to the prediction
of enhanced backscattering [14] and enhanced transmission [15], their
inclusion in the calculation of higher-order moments of the scattered
or transmitted field would also lead to the prediction of new physical
effects.  This expectation was prompted by the results of earlier
theoretical [16,17] and experimental [18-20] investigations of angular
intensity correlation functions in the scattering of classical waves
from volume disordered media.  In a theoretical investigation [9] it
was predicted that three types of correlations occur in such
scattering, viz. short-range correlations, long-range correlations,
and infinite-range correlations.  These were termed the $C^{(1)},
C^{(2)},$ and $C^{(3)}$ correlations, respectively.  The $C^{(1)}$
correlation function includes both the ``memory effect" and the
``reciprocal memory effect" [9,10], so named because of the wave
vector conservation conditions they satisfy.  Both of these effects
have now been observed in volume scattering experiments [16,17].  The
$C^{(2)}$ correlation function has also been observed in volume
scattering experiments [18,19], as has the $C^{(3)}$ correlation
function [20].

Until recently, only the $C^{(1)}$ correlation function had been
studied theoretically and experimentally [1-8].  In a recent series of
papers devoted to theoretical studies of angular correlation functions
of the intensity of light scattered from one--dimensional [9,10] and
two--dimensional [10] randomly rough metal surfaces the long--range
$C^{(2)}$ and infinite--range $C^{(3)}$ correlation functions were
calculated, and two additional types of correlation functions, a
short--range correlation function, named $C^{(10)}$, and a long--range
correlation function, named $C^{(1.5)}$, correlation functions were
predicted.  In very recent experimental work [12] the envelopes of the
$C^{(1)}$ and $C^{(10)}$ correlation functions were measured
experimentally for the scattering of p-polarized light from weakly
rough, one-dimensional gold surfaces. The $C^{(1.5)}, C^{(2)}$, and
$C^{(3)}$ correlation functions have yet to be observed
experimentally.

The question arises as to whether it possible to determine the
relative magnitudes of the different correlation functions from a
knowledge of the experimental parameters of the surface roughness and
its statistical properties.  This question has been raised earlier in
[12,16], but not answered definitively. We therefore address it here
for the case of a one--dimensional random surface defined by the
equation $x_3 =\zx $, on the basis of the single assumption that the
surface profile function $\zx $ is a single-valued function of $x_1$
that constitutes a zero-mean, stationary, Gaussian random process.

The outline of this paper is as follows. In Section 2 we introduce the
angular intensity correlation function and analyze it in terms of the
possible statistics of the scattering matrix.  In Section 3 we
illustrate the conclusions of Section 2 for the simple example of the
scattering of light from the randomly rough surface of a perfect
conductor.  Finally, in Section 4 we present the conclusions drawn
from the results obtained in this work.

\noindent
\setcounter{equation}{0}
\section*{2. The Angular Intensity Correlation Function}
\setcounter{section}{2}

The general angular intensity correlation function $C(q,k|q',k')$ we study
in this work  is
defined by
\bqe
C(q,k|q^{\prime },k^{\prime }) = 
       \langle I(q|k)I(q^{\prime }|k^{\prime})\rangle 
       -\langle I(q|k)\rangle \langle I(q^{\prime }|k^{\prime })\rangle ,
\lb{cin}
\eqe
where the angle brackets denote an average over the ensemble of realizations
of the surface profile function.
The intensity $I(q|k)$ entering this expression is defined in terms of the
scattering matrix $S(q|k)$ for the scattering of light of frequency $\omega $
from a one-dimensional random surface by
\bqe
I(q|k)=\frac {1}{L_1}\left( \frac \omega c\right) |S(q|k)|^2,
\lb{i(q,k)}
\eqe
where $L_1$ is the length of the $x_1$-axis covered by the random
surface, and the wavenumbers $k$ and $q$ are related to the angles of
incidence and scattering, $\to$ and $\ts$, measured counterclockwise
and clockwise from the normal to the mean scattering surface,
respectively, by $k=(\w/c)\sin\to$ and $q=(\w/c)\sin\ts$.  In terms of
the scattering matrix $S(q|k)$ the correlation function $C(q,k|q',k')$
becomes
\bqe
C(q,k|q^{\prime },k^{\prime })=\frac{1}{L_1^2}\frac{\w^2}{c^2}
&&\left [\langle S(q|k)S^*(q|k)S(q^{\prime }|k^{\prime})S^*(q^{\prime 
}|k^{\prime})\rangle\right. \nn\\
&&\quad-\left.\langle S(q|k)S^*(q|k)\rangle \langle S(q^{\prime }|k^{\prime })
S^*(q^{\prime }|k^{\prime })\rangle\right] .
\lb{csm}
\eqe
Since, due to the stationarity of the surface profile function,
$\la S(q|k)\ra $ is diagonal in $q$ and $k$, $\la S(q|k)\ra
=2\pi\delta(q-k)S(k)$, we introduce the incoherent part of the
scattering matrix $\delta S(q|k)=S(q|k)-\la S(q|k)\ra$.  Then, from
the relations between averages of the products of random functions and
the corresponding cumulant averages [21] and omitting all terms
proportional to $2\pi\delta(q-k)$ and/or $2\pi\delta(q'-k')$, as
uninteresting specular effects, \eq{csm} can be rewritten in the form
\bqe
C(q,k|q^{\prime },k^{\prime })=\frac{1}{L_1^2}\frac{\w^2}{c^2}
&&\left [|\langle \delta S(q|k)\delta S^*(q^{\prime }|k^{\prime})\ra|^2 +
|\langle \delta S(q|k)\delta S(q^{\prime }|k^{\prime }\ra|^2\right.\nn\\
&& \quad + \left. \la \delta S(q|k)\delta S^*(q|k)\delta S(q^{\prime }|k^{\prime })
\delta S^*(q^{\prime }|k^{\prime })\ra _c \right ],
\lb{cum}
\eqe
where $\la\cdots\ra _c $ denotes the cumulant average.

Due to the stationarity of the surface profile function $\zeta (x_1)$,
$\langle \delta S(q|k)\delta S^{*}(q^{\prime }|k^{\prime })\rangle $
is proportional to $2\pi \delta (q-k-q^{\prime
}+k^{\prime })$. It gives rise to the contribution to $%
C(q,k|q^{\prime },k^{\prime })$ called $C^{(1)}(q,k|q^{\prime
  },k^{\prime })$, and describes the memory effect and the reciprocal
memory effect.  Similarly, $\langle \delta S(q|k)\delta S(q^{\prime
  }|k^{\prime })\rangle $ is proportional to $2\pi \delta
(q-k+q^{\prime }-k^{\prime })$, and contributes the correlation
function $C^{(10)}(q,k|q^{\prime },k^{\prime })$ to $C(q,k|q^{\prime
  },k^{\prime })$. The third term on the right hand side of \eq{cum}
$\la \delta S(q|k)\delta S^*(q|k)\delta S(q'|k')\delta S^{*}(q^{\prime
  }|k^{\prime })\ra _c $ is proportional to $2\pi\delta (0)= L_1$, due
to the stationarity of the surface profile function $\zeta (x_1)$, and
gives rise to the long--range and infinite--range contributions to
$C(q,k|q^{\prime },k^{\prime })$ given by the sum
$C^{(1.5)}(q,k|q^{\prime },k^{\prime })+ C^{(2)}(q,k|q^{\prime
  },k^{\prime })+C^{(3)}(q,k|q^{\prime }|k^{\prime })$. Thus, we have
separated explicitly the contributions to $C(q,k|q^{\prime },k^{\prime
  })$ that have been named $C^{(1)}(q,k|q^{\prime },k^{\prime })$ and
$C^{(10)}(q,k|q^{\prime },k^{\prime })$.

What is more, from \eq{cum} we can easily estimate the relative
magnitudes of the different contributions to the general correlation
function. Indeed, since $2\pi\delta (0)=L_1,$ when the arguments of
the delta-functions vanish the $C^{(1)}(q,k|q^{\prime },k^{\prime })$
and $C^{(10)}(q,k|q^{\prime },k^{\prime })$ correlation functions are
independent of the length of the surface $L_1$, because they contain
$[2\pi\delta (0)]^2$.  At the same time the remaining term in
\eq{cum}, that yields the sum $C^{(1.5)}(q,k|q^{\prime },k^{\prime
  })+C^{(2)}(q,k|q^{\prime },k^{\prime })+C^{(3)}(q,k|q^{\prime
  }|k^{\prime })$, is inversely proportional to the surface length,
due to the lack of a second delta function. Therefore, in the limit of
a long surface or a large illumination area the long--range and
infinite--range correlations are small compared to short--range
correlation functions, and vanish in the limit of an infinitely long
surface.  Thus, the experimental observation of the $C^{(1.5)}$,
$C^{(2)}$, and $C^{(3)}$ correlation functions requires the use of a
short segment of random surface and/or the use of a beam of narrow
width for the incident field. A detailed discussion of the conditions
under which they may be observed will therefore be deferred to a
separate paper.

The preceding results are consistent with the usual assumptions and
conclusions encountered in conventional speckle theory [22,23]. Thus,
when the surface profile function is assumed to be a stationary random
process, and the random surface is assumed to be infinitely long, the
scattering matrix $S(q|k)$ becomes the sum of a very large number of
independent contributions from different points on the surface. On
invoking the central limit theorem, it is found that $S(q|k)$ obeys
complex Gaussian statistics. In this case \eq{cum} becomes rigorously
\bqe
C(q,k|q^{\prime },k^{\prime })&=&\frac{1}{L_1^2}\frac{\w^2}{c^2}
\left [|\langle \delta S(q|k)\delta S^*(q^{\prime }|k^{\prime})\rangle|^2 +
|\langle \delta S(q|k)\delta S(q^{\prime }|k^{\prime })\rangle |^2\right]\\
&=&C^{(1)}(q,k|q^{\prime },k^{\prime })+C^{(10)}(q,k|q^{\prime },k^{\prime }),
\lb{c1c2}
\eqe
because all cumulant averages of products of more than two Gaussian
random processes vanish. The last term on the right--hand side of
\eq{cum} therefore gives the correction to the prediction of the
central limit theorem due to the finite length of the random surface.

If it is further assumed, as is done in speckle theory, where the
disorder is presumed to be strong, that $\delta S(q|k)$ obeys circular
complex Gaussian statistics [22,23], then $\la \delta S(q|k)\delta
S(q'|k')\ra =0$ and the expression for $C(q,k|q',k')$ simplifies to
\bqe
C(q,k|q^{\prime },k^{\prime })&=&\frac{1}{L_1^2}\frac{\w^2}{c^2}
|\langle \delta S(q|k)\delta S^*(q^{\prime }|k^{\prime})|^2\\
&=&C^{(1)}(q,k|q^{\prime },k^{\prime }).
\eqe
This approximation is often called the factorization approximation
to $C(q,k|q'k')$ [17].

We recall that if the complex random variables $F_1$ and $F_2$ are
jointly circular complex Gaussian random variables, then the conditions
\bqe
\la Re F_1 Re F_2\ra&=&\la Im F_1 Im F_2\ra,\\
\la Re F_1 Im F_2\ra&=&-\la Im F_1 Re F_2\ra,
\lb{ccgs}
\eqe
have to be satisfied.  To analyze how the scattering matrix transforms
from a complex Gaussian random process into a circular complex
Gaussian random process we represent the scattering matrix in the form
$\delta S(q|k)=\delta S_1(q|k)+ i\delta S_2(q|k)$.  The expressions
for the averages of the products of the real and imaginary parts of
$\delta S(q|k)$ can be written in terms of $\la \delta S(q|k)\delta
S^*(q'|k')\ra$ and $\la \delta S(q|k)\delta S(q'|k')\ra$

\bqe
\la \delta S_1(q|k)\delta S_1(q'|k')\ra &=&\frac{1}{2}Re\left[\la \delta S(q|k)
\delta S^*(q'|k')\ra+
\la \delta S(q|k)\delta S(q'|k')\ra\right ]\\
\la \delta S_2(q|k)\delta S_2(q'|k')\ra &=&\frac{1}{2}Re\left[\la \delta S(q|k)
\delta S^*(q'|k')\ra-
\la \delta S(q|k)\delta S(q'|k')\ra\right ]\\
\la \delta S_1(q|k)\delta S_2(q'|k')\ra &=&-\frac{1}{2}Im\left[\la \delta S(q|k)
\delta S^*(q'|k')\ra-
\la \delta S(q|k)\delta S(q'|k')\ra\right ]\\
\la \delta S_2(q|k)\delta S_1(q'|k')\ra &=&\frac{1}{2}Im\left[\la\delta S(q|k)
\delta S^*(q'|k')\ra+
\la \delta S(q|k)\delta S(q'|k')\ra\right ].
\eqe
\lb{rim}

When $q=q'$ and $k=k'$, the average $\la \delta S(q|k)\delta
S(q|k)\ra$, which is proportional to $2\pi \delta(2q-2k)$ due to the
stationarity of the surface profile function, is nonzero only in the
specular direction $q=k$.  Therefore, if the surface is infinitely
long, and if we omit the specular direction, from Eqs. (2.9) --
(2.10), and Eqs. (2.11) -- (2.14) we see that the scattering matrix is
a circular complex Gaussian random process. Consequently, apart from
the specular direction, the speckle contrast $\rho=\sqrt{[\la (\delta
  S(q|k)\delta S^*(q|k) )^2\ra/ \la \delta S(q|k)\delta S^*(q|k)
  \ra^2]-1}$ is unity [22-24].  This result contradicts the
well--known result of Refs. 23 and 24 that the statistics of the
diffuse component of the scattered field is highly non-circular when
the surface is weakly rough, and only in the limit of very rough
surfaces is the circularity of the statistics restored.  The
contradiction stems from the representation of the amplitude of the
scattered field as the convolution of a real--valued amplitude
weighting function and the random phase factor in [23,24]. The
assumption of a real--valued amplitude weighting function, which
represents the finite width of the aperture, is identical to the
assumption of a finite length of the randomly rough surface. As a
result, the statistics of the scattering amplitude is nonstationary in
[23,24].  In the present work we are interested only in the case where
the statistics of the surface profile function, as well as of the
scattering matrix, is stationary.
  
The set of the scattering matrices $\delta S(q|k)$ is a set of jointly
circular complex Gaussian random variables when $\la \delta
S(q|k)\delta S(q'|k')\ra$ vanishes. But when $\la \delta S(q|k)\delta
S(q'|k')\ra$ vanishes the correlation function $C^{(10)}$ vanishes,
since, within a coefficient, $C^{(10)}(q,k|q'k')\sim|\la \delta
S(q|k)\delta S(q'|k')\ra|^2$.

Thus, calculations and measurements of the correlation function
$C(q,k|q',k')$ yields important information about the statistical
properties of the amplitude of the scattered field. If the random
surface is such that only the $C^{(1)}$ and $C^{(10)}$ correlation
functions are observed, then $S(q|k)$ obeys complex Gaussian
statistics. If the random surface is such that only $C^{(1)}$ is
observed, then $S(q|k)$ obeys circular complex Gaussian statistics.
Finally, if the random surface is such that $C^{(1.5)}$, $C^{(2)}$ and
$C^{(3)}$ are observed in addition to both $C^{(1)}$ and $C^{(10)}$,
then $S(q|k)$ is not a Gaussian random process, but the statistics it
obeys in this case are not known at the present time.

To conclude this section we introduce the the normalized angular
intensity correlation functions of interest to us, which in terms of
$\delta S(q|k)$ are defined by
\bqe
\Xi^{(1)} (q,k|q^{\prime },k^{\prime })=
\frac{|\langle \delta S(q|k )\delta S^*(q'|k')\ra|^2}
{\langle \delta S(q|k)\delta S^*(q|k)\rangle \langle \delta S(q^{\prime }|k^{\prime })\delta S^*(q^{\prime }|k^{\prime })\rangle },
\lb{norm1}
\eqe
and
\bqe
\Xi^{(10)} (q,k|q^{\prime },k^{\prime })=
\frac{|\langle \delta S(q|k )\delta S(q'|k')\ra|^2}
{\langle \delta S(q|k)\delta S^*(q|k)\rangle \langle \delta S(q^{\prime }|k^{\prime })\delta S^*(q^{\prime }|k^{\prime })\rangle }.
\lb{norm10}
\eqe
We introduce also the the envelopes $C^{(1)}_0$  and 
$C^{(10)}_0$ of the correlation functions
$C^{(1)}$  and $C^{(10)}$ , which we define by
\bqe
C^{(1)}(q,k|q',k')=2\pi\delta(q-k-q'+k')C^{(1)}_0(q,k|q',q'-q+k)
\eqe
and
\bqe
C^{(10)}(q,k|q',k')=2\pi\delta(q-k+q'-k')C^{(10)}_0(q,k|q'q'+q-k).
\eqe
\setcounter{equation}{0}
\setcounter{section}{3}
\section*{3. Light scattering from a perfectly conducting randomly rough
surface in the framework of phase perturbation theory. }

In this Section we study the statistical properties of the scattering
matrix for the problem of the scattering of a scalar plane wave from a
randomly rough infinitely long surface defined by the equation
$x_3=\zx $.  The region $x_3>\zx $ is vacuum, while the region
$x_3<\zx $ is a perfectly conducting medium.  It is assumed that the
Dirichlet boundary condition is satisfied on the surface $x_3=\zx $.

The surface profile function $\zeta (x_1)$ is assumed to be a
single-valued function of $x_1$ that is differentiable and  constitutes
a zero-mean, stationary, Gaussian random process defined by the properties 
\sctr{0}
\begin{eqnarray}
\langle \zeta (x_1)\rangle  =0, \,\, \,\,\,\,\langle \zeta (x_1)\zeta
(x_1^{\prime })\rangle  &=&\delta ^2W(|x_1-x_1^{\prime }|).
\lb{zeta}
\end{eqnarray}
In Eqs.\ef{zeta} the angle brackets denote an average over the
ensemble of
realizations of $\zeta (x_1)$, and $\delta =\langle \zeta ^2(x_1)\rangle ^{%
  \frac 12}$ is the rms height of the surface, $W(|x_1|)$ is the
surface height autocorrelation function.  In numerical examples we
will use the Gaussian form for $W(|x_1|)$
\sctr{0}
\begin{eqnarray}
W(|x_1|)=\exp (-x_1^2/a^2),
\lb{w(x)}
\end{eqnarray}
where $a$ is the transverse correlation length of
the surface roughness.

A reciprocal phase--perturbation theory for the
scattering matrix $S(q|k)$ was constructed in  Refs. 25 and 26.
The term of lowest order in the surface profile function
was shown to have the form
\bqe
S(q|k)=\iii dx_1e^{-i(q-k)x_1}e^{-2i\sqrt{\aoq\aok}\zx}.
\lb{ppt}
\eqe
Since
\bqe
\la S(q|k)\ra =2\pi\delta(q-k)e^{-2\delta^2\aoq\aok},
\eqe
we can write the expression for $\delta S(q|k)$ as
\bqe
\delta S(q|k)=\iii dx_1e^{-i(q-k)x_1}\left [e^{-2i\sqrt{\aoq\aok}\zx}-
e^{-2\delta^2\aoq\aok}\right ].
\lb{dpt}
\eqe

We calculate the averages $\la \delta S(q|k)\delta 
S^*(q'|k')\ra$ and $\la \delta S(q|k)\delta S(q'|k')\ra$  using the 
expression \ef{dpt} for the scattering
matrix.
For $\la \delta S(q|k)\delta S^*(q'|k')\ra$ we obtain
\bqe
\la \delta S(q|k)\delta S^*(q'|k')\ra &=&\iii dx_1\iii dx_1'
e^{-i(q-k)x_1+i(q'-k')x_1^{\prime}}\nn\\
&&  \quad \times 
     \left< \left[
         e^{-2i\sqrt{\aoq\aok}\zx}-e^{-2\delta^2\aoq\aok}\right]
     \right. \nn\\
&&  \quad \times \left. \left[ e^{2i\sqrt{\aoqp \aokp} \zxp}-
     e^{-2\delta^2\aoqp\aokp}\right ]\right> \\
& = & e^{-2\delta^2(\aoq\aok+\aoqp\aokp)}\iii dx_1\iii dx_1'
     e^{-i(q-k)x_1+i(q'-k')x_1^{\prime}}\nn\\
& & \quad \times \left [e^{4\delta^2\sqrt{\aoq\aoqp\aok\aokp}
W(|x_1-x_1'|)}-1\right ]\\
&=& 2\pi\delta(q-k-q'+k')e^{-2\delta^2(\aoq\aok+\aoqp\aokp)}\nn\\
& & \quad \times \iii du
\left[e^{4\delta^2\sqrt{\aoq\aoqp\aok\aokp}W(|u|)}-1\right ]e^{-i(q'-k')u},
\lb{s1}
\eqe
while for  $\la \delta S(q|k)\delta S(q'|k')\ra$  we have
\bqe
\la \delta S(q|k)\delta S(q'|k')\ra &= &\iii dx_1\iii 
dx_1'e^{-i(q-k)x_1-i(q'-k')x_1^{\prime}}\nn\\
& & \quad \times  \left< 
      \left[e^{-2i\sqrt{\aoq\aok}\zx}-e^{-2\delta^2\aoq\aok}\right]
    \right. \nn\\
& & \quad \times \left.\left[e^{2i\sqrt{\aoqp \aokp}
        \zxp}-e^{-2\delta^2\aoqp\aokp}\right] \right> \\
&= & e^{-\delta^2(\aoq\aok+\aoqp\aokp)/2}\iii dx_1\iii dx_1'
       e^{-i(q-k)x_1-i(q'-k')x_1^{\prime}}\nn\\
& & \quad \times \left[e^{-4\delta^2\sqrt{\aoq\aoqp\aok\aokp}
       W(|x_1-x_1'|)}-1\right]\\
&=&2\pi\delta(q-k+q'-k')e^{-2\delta^2(\aoq\aok+\aoqp\aokp)}\nn\\
&& \quad \times \iii du
\left [e^{-4\delta^2\sqrt{\aoq\aoqp\aok\aokp}W(|u|)}-1\right]e^{-i(q'-k')u}.
\lb{s10}
\eqe

It is readily seen that in contrast to $\la \delta S(q|k)\delta
S^*(q'|k')\ra$ the average $\la \delta S(q|k) \delta S(q'|k')\ra$
vanishes with increasing roughness parameters $\delta$ and $a$, due to
the negative exponential under the integral sign in the last line of
\eq{s10}.  Plots of the normalized correlation functions $\Xi
^{(1)}(q,k|q^{\prime },k^{\prime })$ and $ \Xi ^{(10)}(q,k|q^{\prime
  },k^{\prime })$ as functions of $\delta$ for different values of $a$
are presented in Fig. 1 (a), while plots of the envelopes of the
correlation functions $C^{(1)}$ and $C^{(10)}$ as functions of
$\delta$ for different values of $a$ are presented in Fig. 1(b), for
fixed values of $q$, $k$ and $q'$, while $k'$ is determined by the
constraint of the corresponding $\delta-$function.  When calculating
the results presented in Figs. 1 (a) and (b) the value of $q'$ was
chosen to produce the same values of $C^{(1)}$ and $C^{(10)}$ in the
limit of a weakly rough surface.  From the plots presented in Fig.
1(a) we see that $\Xi ^{(10)}(q,k|q^{\prime },k^{\prime })$ vanishes
even for quite moderately weakly rough surfaces for which $\Xi
^{(1)}(q,k|q^{\prime },k^{\prime })$ is still about unity.  We note
that $C^{(1)}$ also decreases with increasing $\delta$ (Fig. 1 (b)).
Using Eqs.(2.11) - (2.14), \ef{s1} and \ef{s10} we obtain the
expressions for $\la (\delta S_1(q|k))^2\ra$, $\la (\delta
S_2(q|k))^2\ra$ and $\la \delta S_1(q|k)\delta S_2(q|k)\ra$
\bqe
\la (\delta S_1(q|k))^2\ra & =&e^{-4\delta^2\aoq\aok}\left[\frac{L_1}{2}
\iii du \cos(q-k)u \left (e^{\delta^2\aoq \aok W(|u|)} -1\right)\right.\nn\\
&& \quad + \left.\frac{1}{2}\pi\delta(q-k) \iii  \cos(q-k)u\left ( e^{-\delta^2
\aoq \aok W(|u|)}-1\right )\right ],
\eqe
and

\begin{figure}
\begin{center}
  \begin{tabular}{@{}c@{\hspace{1.0cm}}c@{}}
    \epsfig{figure=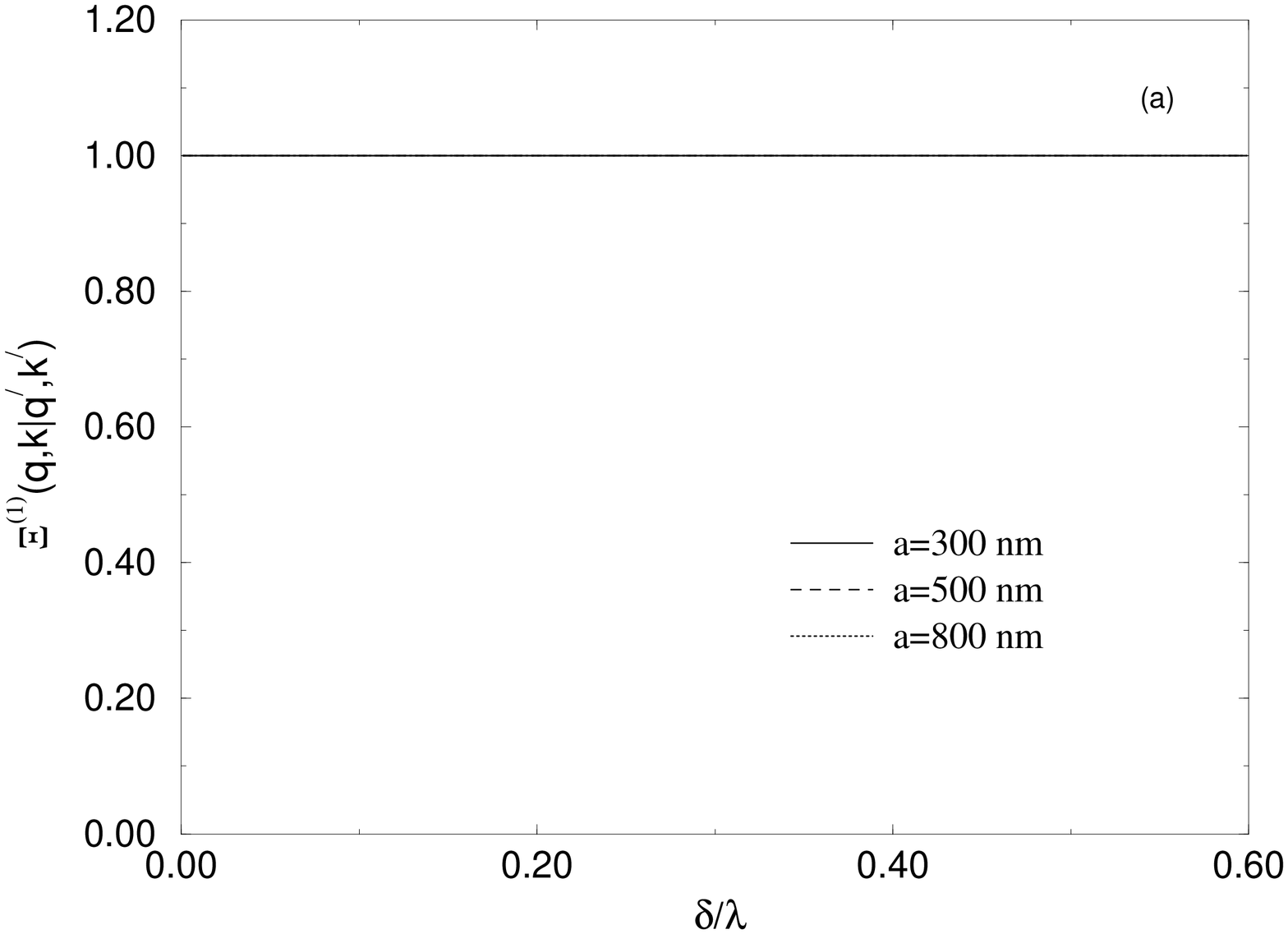,height=3in,width=3in,angle=-90} &
    \epsfig{figure=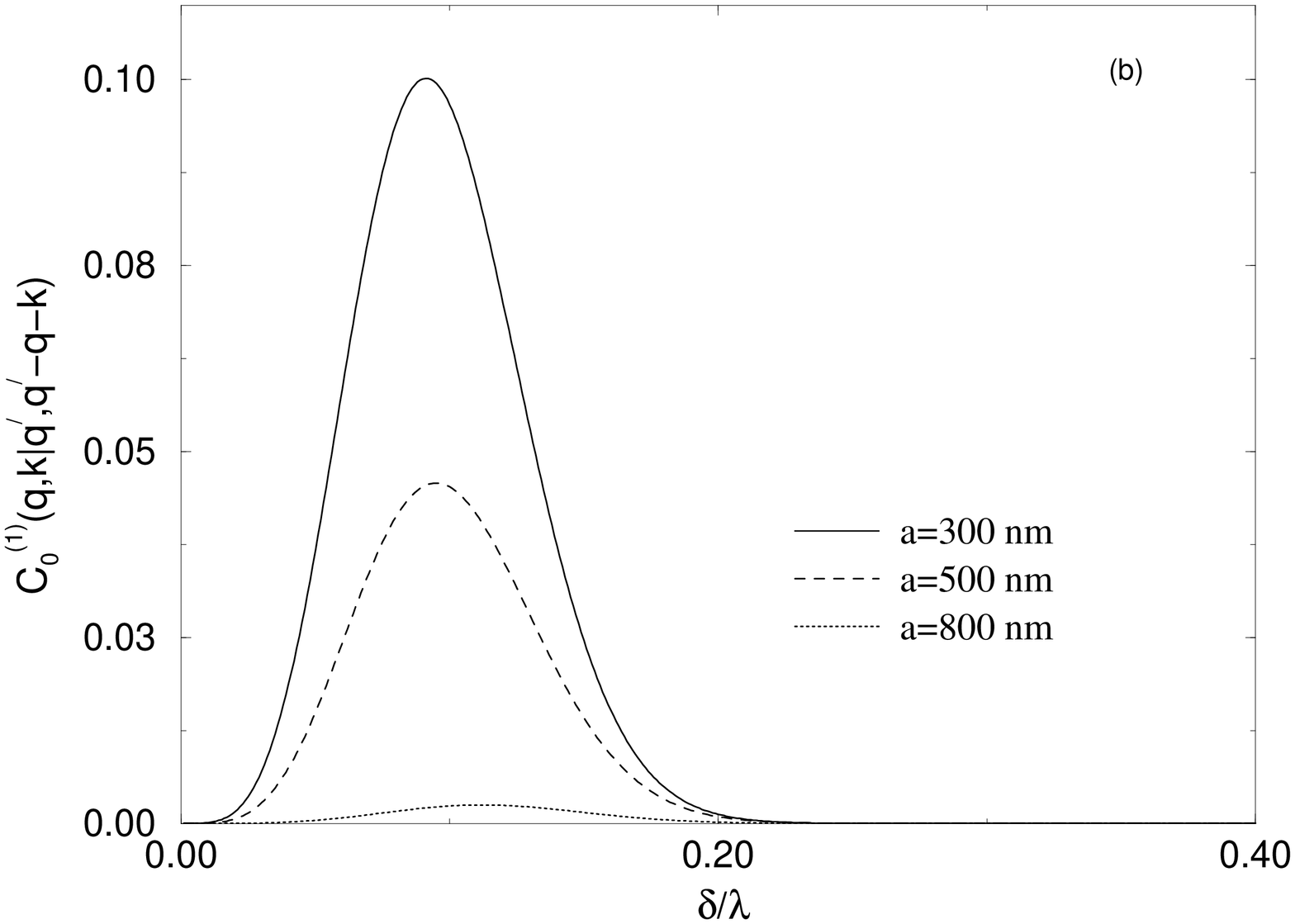,height=3in,width=3in,angle=-90}
    \\
    \epsfig{figure=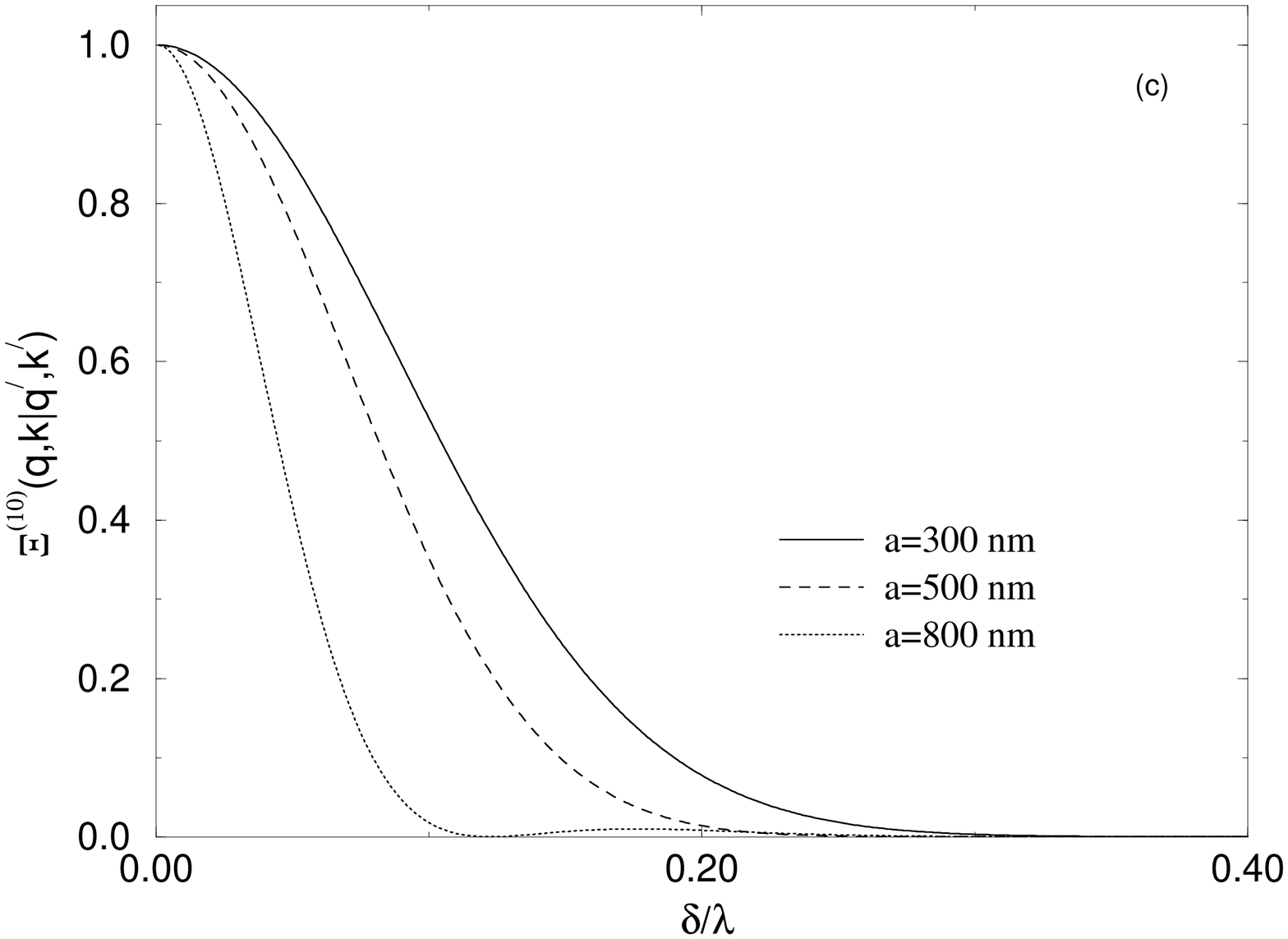,height=3in,width=3in,angle=-90} &
    \epsfig{figure=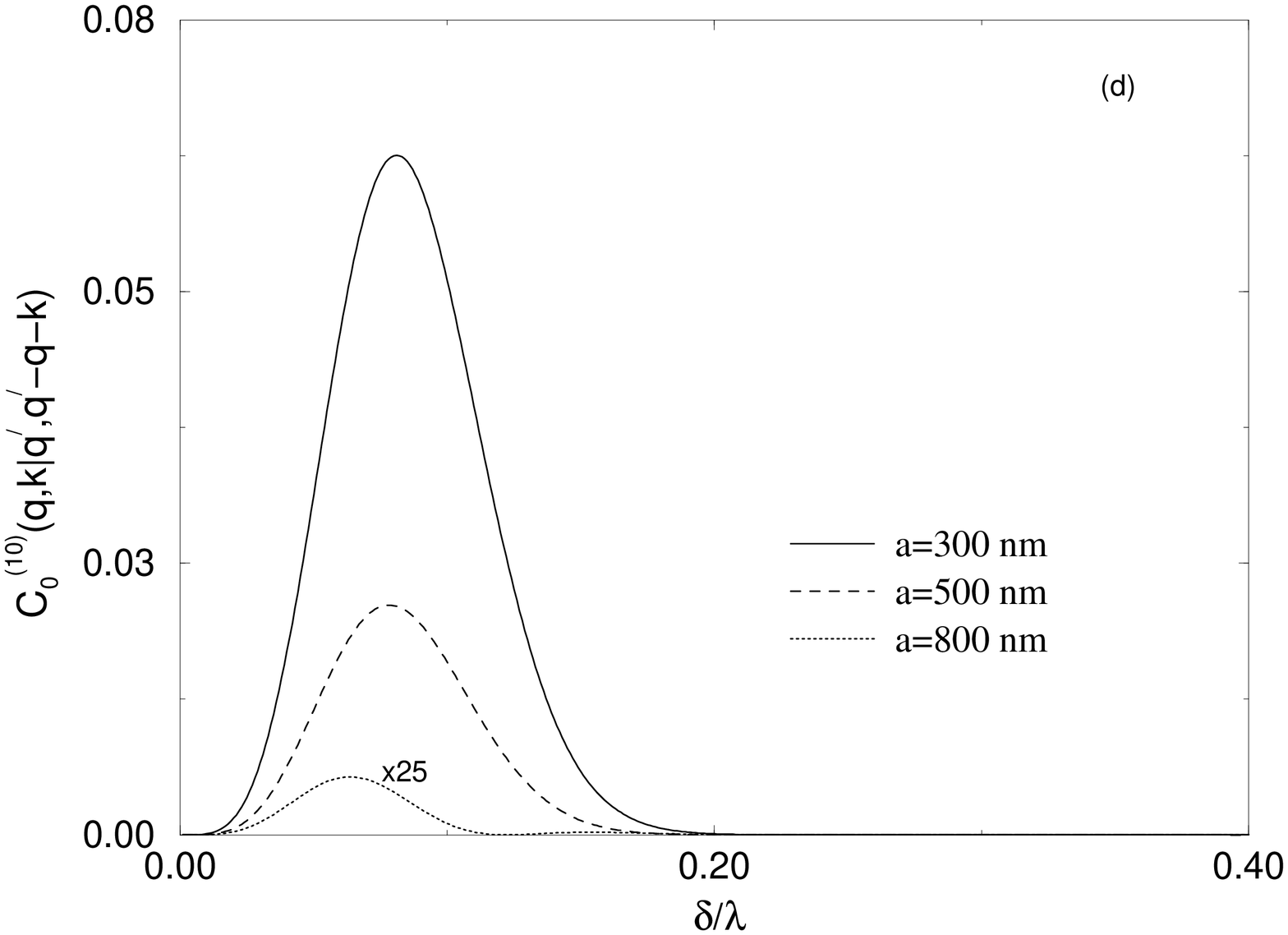,height=3in,width=3in,angle=-90}
  \end{tabular}
\end{center} 
\caption{The normalized correlation functions $\Xi^{(1)}$ (a) and
  $\Xi^{(10)}$ (c), and the envelopes $C^{(10)}_0$ (b) and
  $C^{(10)}_0$ (d) as functions of $\delta/\lambda$ for values of the
  transverse correlation length $a=300~\mbox{nm}$, $500~\mbox{nm}$,
  and $800~\mbox{nm}$. The incident light was $s-$polarized and of
  wavelength $632.8 nm$. The scattering medium was a randomly rough
  perfect conductor.  Furthermore $\to =30^{\circ}$, $\ts
  =0^{\circ}$, and $\theta'_s=0^{\circ}$. In Fig. 1a the results
  for the different correlation lengths considered could not be
  distinguished.}
\end{figure}
\vspace*{0.2cm}

\noindent
\bqe
\la (\delta S_2(q|k))^2\ra & =&e^{-4\delta^2\aoq\aok}\left[\frac{L_1}{2}
\iii du \cos(q-k)u \left (e^{\delta^2\aoq \aok W(|u|)} -1\right )\right.\nn\\
&& \quad - \left. \frac{1}{2}\pi\delta(q-k) \iii  \cos(q-k)u
\left (e^{-\delta^2\aoq \aok W(|u|)}-1\right )\right],
\eqe
while
\bqe
\la \delta S_1(k|k)\delta S_2(k|k)\ra  =0.
\eqe

In Fig. 2 we present plots of the ratio $\la (\delta
S_2(k|k))^2\ra/\la (\delta S_1(k|k))^2\ra $ as a function of the rms
height of the surface roughness $\delta$. Since this ratio is
calculated for the specular direction $q=k$, it is independent of the
transverse correlation length $a$. From the plot presented it is
easily seen that for large values of the rms height the incoherent
part of the scattering matrix, $\delta S(q|k)$, becomes a circular
complex Gaussian variable, even in the specular direction.
\begin{figure}
\begin{center}
\epsfig{figure=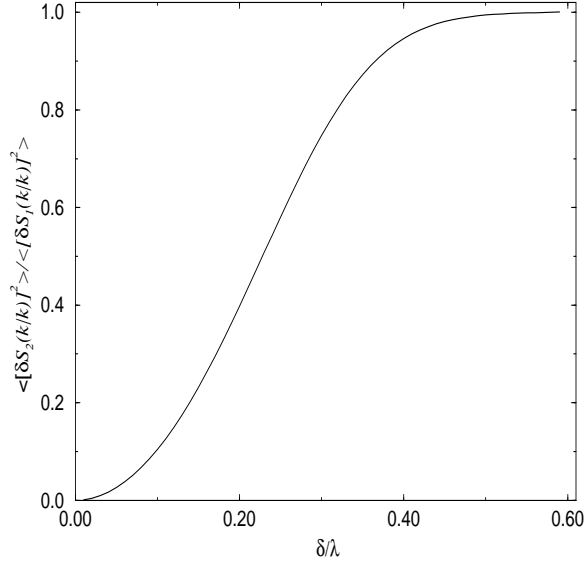,height=3in,width=3.in,angle=-90}
\end{center}
\caption{The ratio $\la (\delta S_2(k|k))^2\ra/\la (\delta S_1(k|k))^2\ra$
as a function of $\delta/\lambda $.}
\end{figure}

\noindent
\setcounter{equation}{0}
\setcounter{section}{4}
\section*{4. Light scattering from a randomly rough penetrable
surface}

The results of the preceding Section enable us to make several conclusions
when studying the scattering of light from a randomly rough  surface of
a penetrable medium.  For simplicity we consider here the scattering
of $s-$polarized light
from randomly rough surface of a  medium characterized
by a  dielectric function $\e (\w)$.
As is well known (see, e.g. [27-29]) if the surface profile
function is such that the conditions for the applicability of the
Rayleigh hypotesis are satisfied the scattering amplitude $R(q|k)$
obeys the reduced Rayleigh equation. Rewritten in terms of  the scattering
matrix  $S(q|k)$ it has the form
\bqe
S(q|k)=2\pi\delta(q-k)R_0(k)+N(q|k) + \iii \frac{dp}{2\pi}M(q|p)S(p|k),
\lb{rre}
\eqe
where  for the case of the scattering of  $s-$polarized light,
\bqe
R_0(k)=\frac{\aok -\ak}{\aok+\ak},
\eqe
\bqe
\aok=\sqrt{\frac{\w^2}{c^2} -k^2}, \, \,\,\,\,\,\,\,\,
\ak=\sqrt{\ew\frac{\w^2}{c^2}-k^2},
\eqe
\bqe
N(q|k)=-(\e-1)\frac{\w^2/c^2}{\aoq+\aq}\sqrt{\frac{\aoq}{\aok}}
\frac{J(\alp + \aok |p-k)}{\alp + \aok},
\eqe
\bqe
M(q|k)= -(\e-1))\frac{\w^2/c^2}{\aoq+\aq}\sqrt{\frac{\aoq}{\alpha_0(p)}}
\frac{J(\alp - \aok |p-k)}{\alp - \aok} ,
\eqe
and
\bqe
J(\gamma|Q)=\iii dx_1 e^{-iQx_1}\left (e^{-i\gamma\zx }-1\right ).
\eqe

\begin{figure}
\begin{center}
  \begin{tabular}{@{}c@{\hspace{1.0cm}}c@{}}
    \epsfig{figure=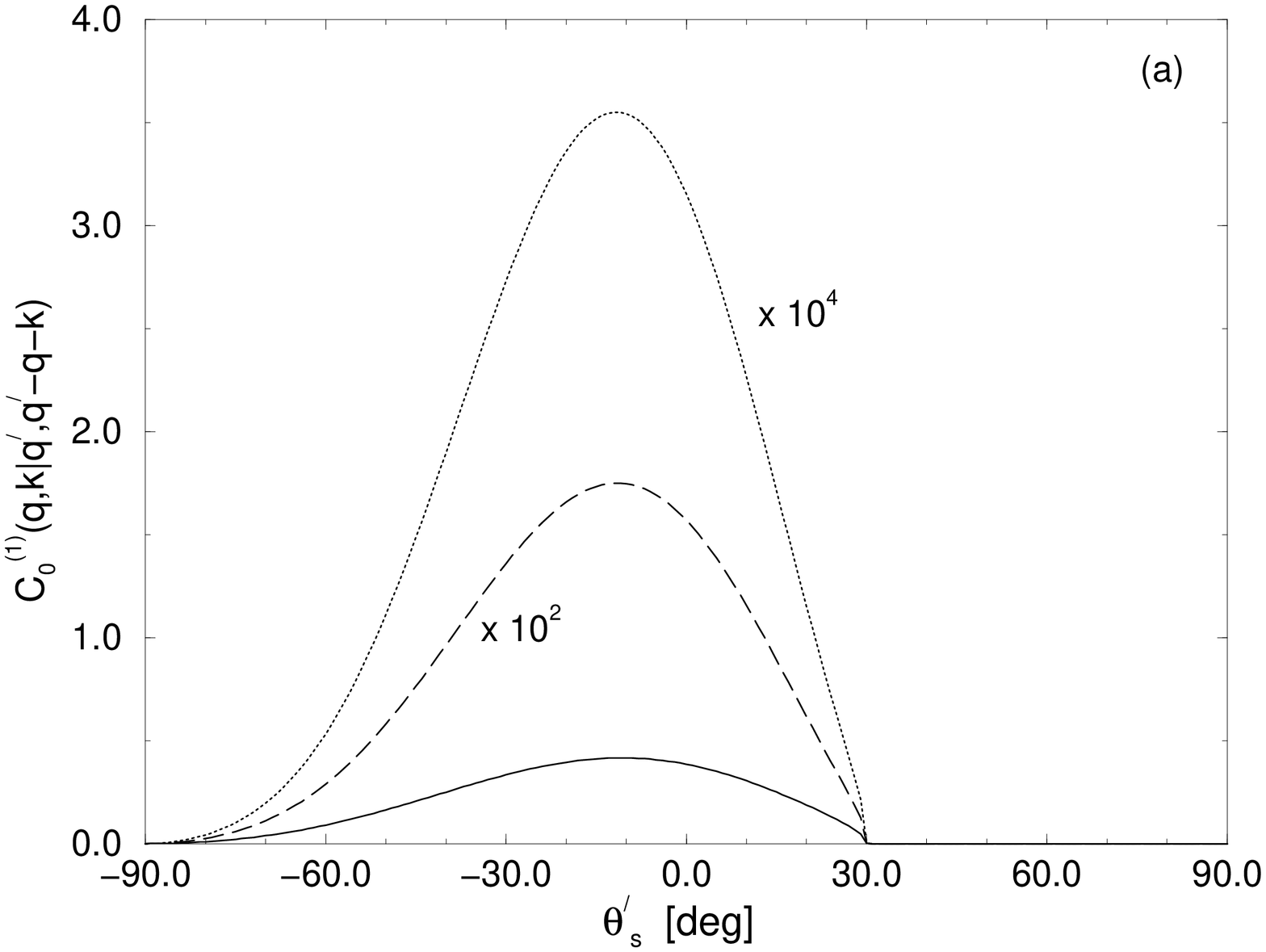,height=2.5in,width=3in,angle=-90} &
    \epsfig{figure=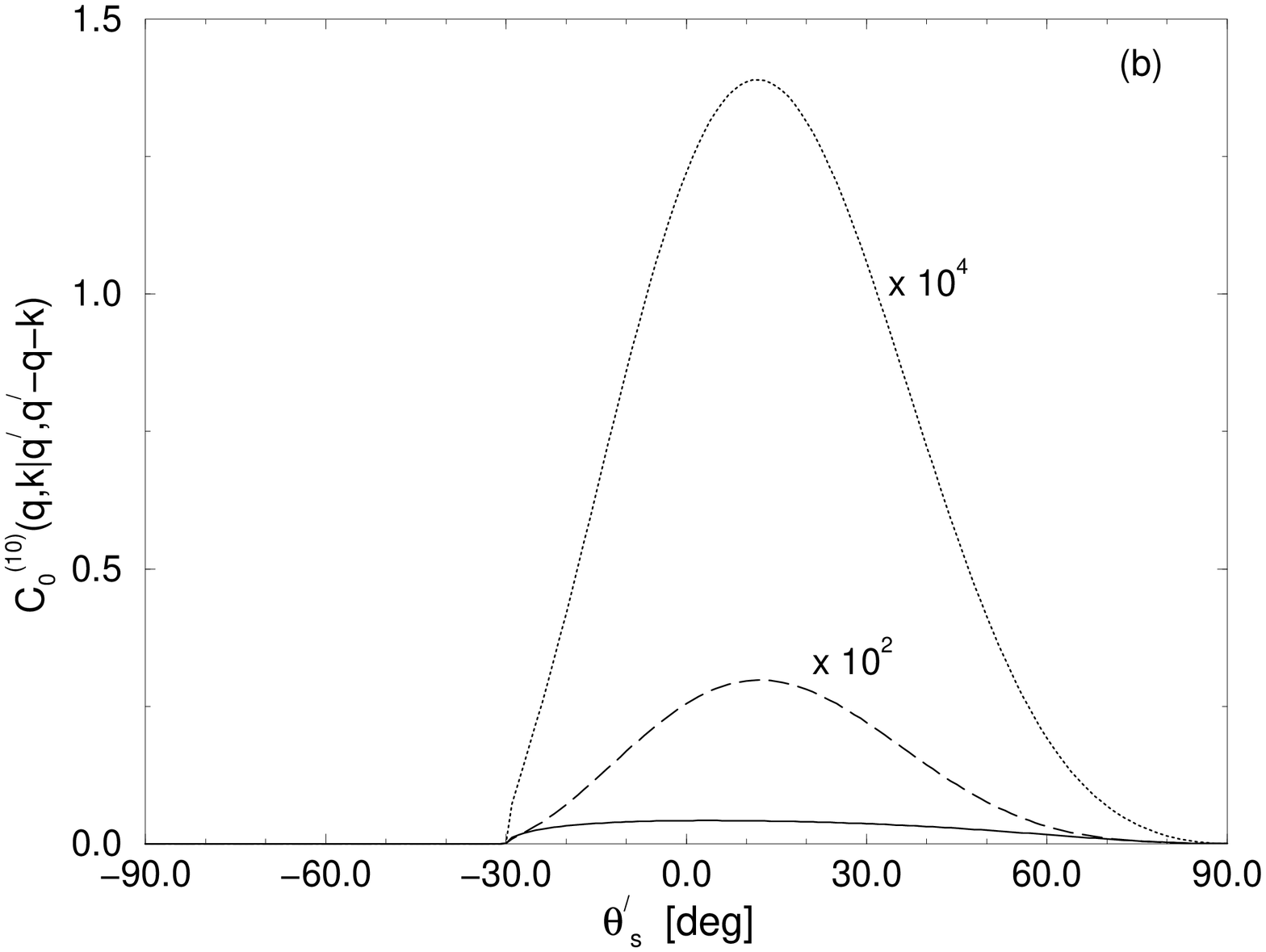,height=2.5in,width=3in,angle=-90}
  \end{tabular}
\end{center}
\caption{ The envelopes of the $C^{(1)}$ (a) and $C^{(10)}$
  (b) correlation functions as functions of $\theta_s^{\prime}$ for
  $\to =30^{\circ}$ and $\ts =0^{\circ}$, while $\theta'_0$ is
  determined by the constraints of the corresponding $\delta-$
  functions, for the scattering of $s-$polarized light from a randomly
  rough silver surface with $a=500 nm $ and $\delta =20\, nm$ (solid
  lines), $\delta =50\, nm $ (dashed lines), and $\delta = 100\, nm
  $ (dotted lines).}
\end{figure}
\vspace*{0.8cm}

\noindent
We can  write the solution of \eq{rre} formally as
\bqe
S(q|k)&=& R_0(k)2\pi\delta(q-k)+F(q|k)+\iii \frac{dp}{2\pi}M(q|p)F(p|k)+\nn\\
&& \quad +\iii \frac{dp}{2\pi} M(q|p)
\iii \frac{dp'}{2\pi} M(p|p')F(p'|k)+\cdots,
\lb{iter}
\eqe
where
\bqe
F(q|k)=N(q|k)+M(q|k)R_0(k),
\eqe
and we
keep all terms in the infinite iterative series.
Both $N(q|k)$ and $M(q|p)$ contain the surface disorder only in the functions
$J(\gamma|Q)$. Therefore, having in hand the recipe for calculating
the average of the product of any number of  functions $J(\gamma|Q)$,
we can calculate, in principle, both $\la \delta S(q|k)\delta S(q'|k')\ra $
and $\la \delta S(q|k)\delta S^*(q'|k')\ra $. The basics of such calculations
were described in Ref. [30].

To calculate the averages $\la \delta  S(q|k) \delta S(q'|k')\ra $
and $\la \delta S(q|k) \delta S^*(q'|k')\ra $ we multiply the series 
(4.7) for $S(q|k)$ by the corresponding series for $S(q'|k')$, and 
average the product term-by-term.  From the result we subtract the 
product $\la S(q|k)\ra \la S(q'|k')\ra$.  In a similar fashion we 
calculate the average $\la \delta S(q|k)\delta S^*(q'|k')\ra$ by 
multiplying the series (4.7) for $S(q|k)$ by the complex conjugate 
of the corresponding series for $S(q'|k')$, averaging the product 
term-by-term, and subtracting the product $\la S(q|k)\ra\la 
S^*(q'|k')\ra$ from the result.  In the product$\la \delta 
S(q|k)\delta S^*(q'|k')\ra$ the contribution of n$^{\rm th}$ order 
in the functions $J(\gamma |Q)$ and $J^*(\gamma |Q)$ contains $n-1$ 
terms of the form
\bqe
\sum^{n-1}_{m=1} \left\{\left\la \prod^{m}_{r=1} 
J(\gamma_r|Q_r)\prod^{n-m}_{s=1} J^*(\gamma'_s|Q'_s)\right\ra  
- \left \la \prod^{m}_{r=1} J(\gamma_r|Q_r)\right \ra \left\la 
\prod^{n-m}_{s=1} J^*(\gamma '_s|Q'_s)\right \ra \right\} .
\eqe
To obtain a nonzero contribution, for each value of $m$ at least one 
$J(\gamma_r|Q_r)$ must be contracted with at least one $J^*(\gamma 
'_s|Q'_s)$.  
Therefore, each term in this sum contains at 
least one factor with a postive exponential of the form $\exp \{ 
\delta^2\gamma \gamma 'W(|u|)\} - 1$.  In contrast, when calculating 
$\la \delta S (q|k)\delta S(q'|k')\ra$ the contribution of the 
n$^{\rm th}$ order  in the functions $J(\gamma |Q)$ contains the sum
\bqe
\sum^{n-1}_{m=1} \left\{ \left\la \prod^{m}_{r=1} 
J(\gamma_r|Q_r)\prod^{n-m}_{s=1}J(\gamma '_s|Q'_s)\right\ra - 
\left\la 
\prod^{m}_{r=1} J(\gamma_r|Q_r)\right \ra \left \la \prod^{n-m}_{s=1}J(\gamma 
'_s|Q'_s)\right \ra \right\} .
\eqe

\begin{figure}
\begin{center}
  \begin{tabular}{@{}c@{\hspace{1.0cm}}c@{}}
    \epsfig{figure=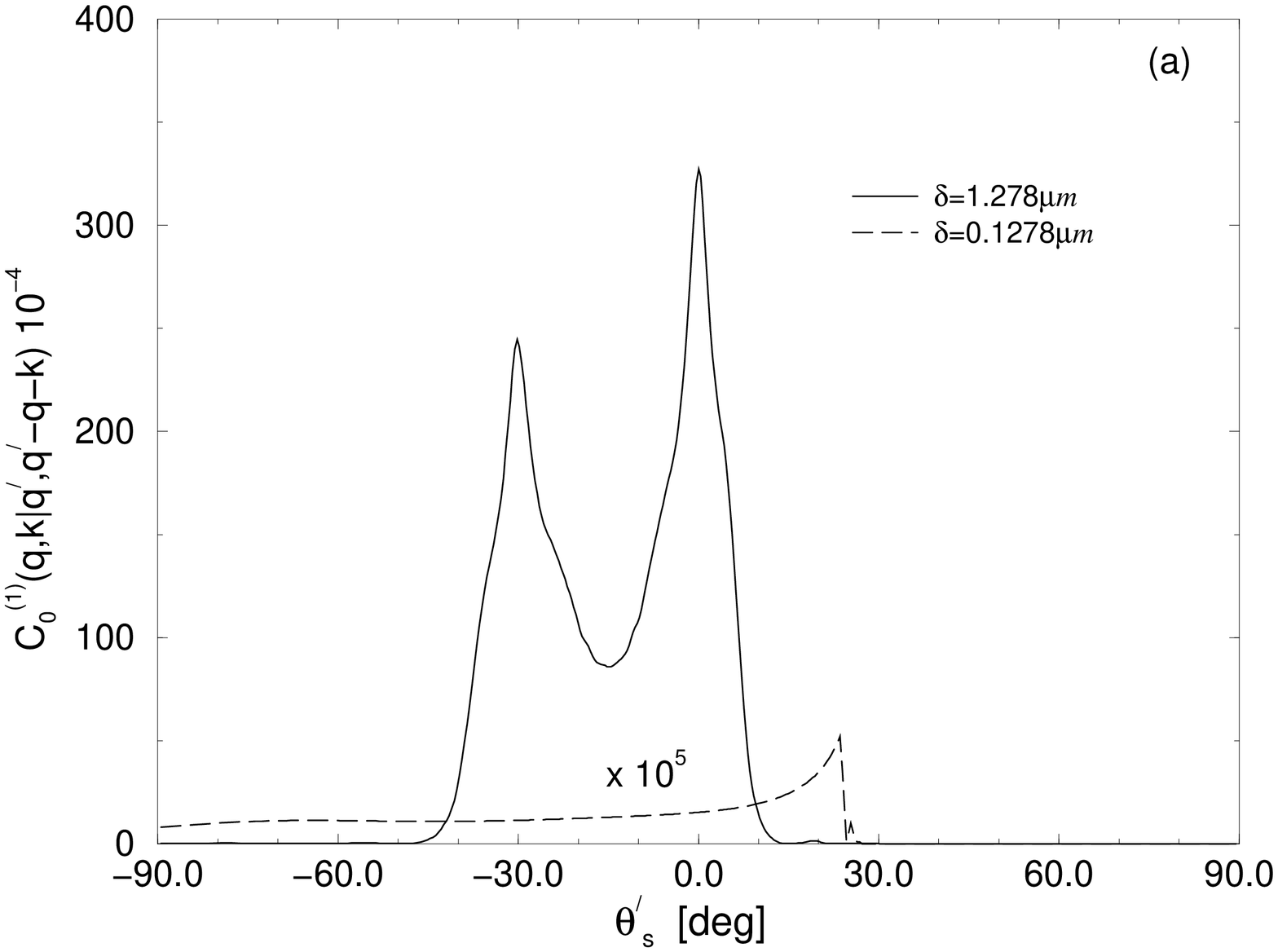,height=2.5in,width=3in,angle=-90} &
    \epsfig{figure=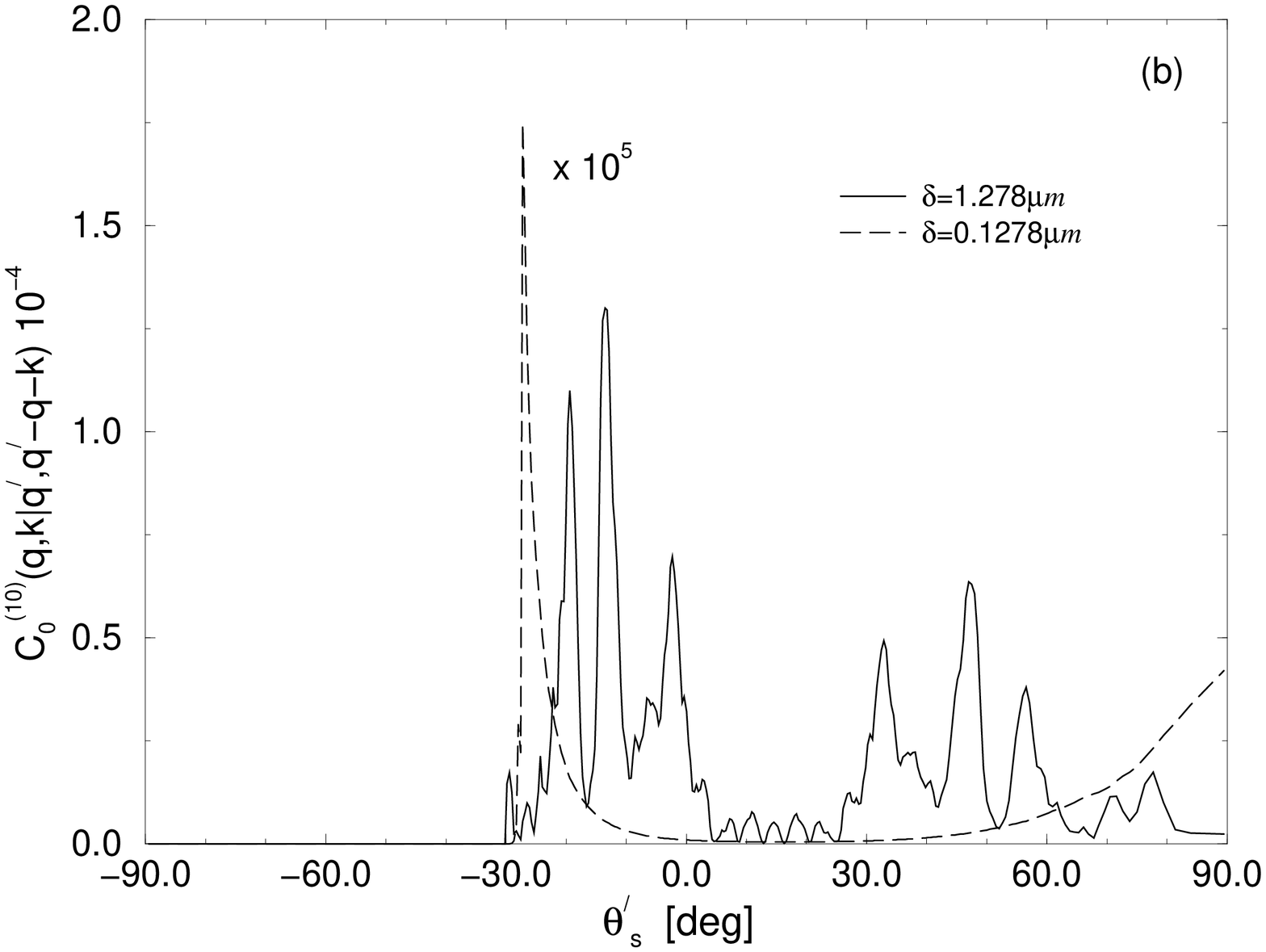,height=2.5in,width=3in,angle=-90}
  \end{tabular}
\end{center}
\caption{  The same as in Fig. 3, but for $a = 3.85 \mu m$ and
$\delta = 1.278 \mu m$ (solid lines) and $\delta = 0.1278 \mu m$ (dashed lines).}
\end{figure}

\vspace*{0.3cm}

In this case, to obtain a nonzero contribution, for each value of $m$
at least one $J(\gamma_r|Q_s)$ must be contracted with at least one
$J(\gamma '_s|Q'_s)$.  Therefore, each term in this sum contains only
negative exponentials of the form $\exp \{ - \delta^2\gamma\gamma
'W(|u|)\} -1$.  Owing to this lack of the positive exponential, $\la
\delta S(q|k)\delta S(q'|k')\ra $ vanishes when the roughness
parameters increase.

In Fig. 3 we present plots of the envelopes $C^{(1)}_0$ and
$C^{(10)}_0$ of the correlation functions $C^{(1)}$ (Fig. 3(a)) and
$C^{(10)}$ (Fig. 3(b)) as functions of $\theta_s^{\prime}$ for fixed
values of $\to$ and $\ts$, while $\theta_0^{\prime}$ is determined by
the constraints of the corresponding $\delta-$functions.  The
calculations were carried out for the scattering of $s-$polarized
light, of $612.7 nm $ wavelength, from a weakly rough random surface
of a silver characterized by the complex dielectric constant $\e=-17.2
+i0.479$ for different values of the roughness parameters $\delta$ and
$a$.  In calculating the results presented in Fig. 3 we kept all terms
in the infinite iterative series \eq{iter} which would give the
contributions to the averages we calculate through terms of
$O(\delta^8$) if they were to be expanded in powers of the small
parameter $ (\w/c)\delta$.

In Fig.~4 we present rigorous numerical simulation calculation
results~\cite{AnnPhys} for the envelopes of the correlation functions
$C^{(1)}$ (Fig.~4a) and $C^{(10)}$~(Fig.~4b). The surface parameters
used here were the same as those used in obtaining Fig.~3 except that
the roughness now was $\delta=1.278\,\mu m$~(solid lines) and
$\delta=0.1278\,\mu m$~(dashed lines).  It should be pointed out that
for the scattering of $s$-polarized light from a weakly rough random
metal surface there should be no memory- or reciprocal memory-effect
present in $C^{(1)}_0$.  This is indeed confirmed by our numerical
calculations where the $C^{(1)}_0$ for $\delta=0.1278\,\mu m$~(Fig.
4a, dashed line) is a smooth function of its argument, as well as by
the results presented in Fig. 3a.  In particular, there are no peaks
at angles $\theta=0^\circ$ and $30^\circ$, which are the positions of
the memory- and reciprocal memory-effects. As the roughness is
increased to $\delta=1.278\,\mu m$ one sees from Fig.~4a (solid line)
that the overall amplitude of the envelope $C^{(1)}_0$ is increased
and, more important, that two peaks have developed at the
aforementioned angles. These peaks are due, in the large roughness
limit, to volume waves scattered multiply at the rough surface. In
Fig.~4b the corresponding results for the $C^{(10)}_0$-envelopes are
presented. It is observed that in the low roughness limit this
envelope is structureless, and that $C^{(1)}_0$ and $C^{(10)}_0$ are
roughly of the same order of magnitude. However, as $\delta$ is
increased, the scattering matrix $S(q|k)$ starts to obey circular
complex Gaussian statistics, and thus as discussed earlier, the
envelope $C^{(10)}_0$ should in principle vanish. From our numerical
results for $\delta=1.278\,\mu m$~(solid line) we indeed see that
$C^{(10)}_0$ is much smaller then the corresponding $C^{(1)}_0$ shown
in Fig.~4a.  In fact $C^{(10)}_0$ is just noise, consistent with this
function vanishing in the large roughness limit.

\noindent
\setcounter{equation}{0}
\setcounter{section}{5}
\section*{5. Conclusions  }

In this paper we calculated the angular intensity correlation
functions by means of an approach that explicitly separates out
different contributions to it.  We have shown that calculations and
measurements of the correlation function $C(q,k|q',k')$ yields
important information about the statistical properties of the
amplitude of the scattered field. In particular, we have shown that
the short--range correlation function $C^{(10)}$ is, in a sense, a
measure of the noncircularity of the complex Gaussian statistics of
the scattering matrix.  Thus, if the random surface is such that only
the $C^{(1)}$ and $C^{(10)}$ correlation functions are observed, then
$S(q|k)$ obeys complex Gaussian statistics. If the random surface is
such that only $C^{(1)}$ is observed, then $S(q|k)$ obeys circular
complex Gaussian statistics.  Finally, if the random surface is such
that $C^{(1.5)}$, $C^{(2)}$ and $C^{(3)}$ are observed in addition to
both $C^{(1)}$ and $C^{(10)}$, then $S(q|k)$ is not a Gaussian random
process.

\setcounter{equation}{0}
\setcounter{section}{5}

\acknowledgements 

The work of T .A. L. and A. A. M.was supported in part by Army
Research Office Grant No. DAAD 19--99--1--0321.  I. S. would like to
thank the Research Council of Norway (Contract No. 32690/213) and
Norsk Hydro ASA for financial support.

\end{document}